\documentclass[9pt,twocolumn,twoside]{opticajnl}
\journal{opticajournal} 
\usepackage{braket}

\usepackage{lineno}

\title{Capturing the spectrotemporal structure of a biphoton wave packet with delay-line-anode single-photon imagers}

\author[1]{Ozora Iso}
\author[2]{Kensuke Miyajima}
\author[1]{Ryosuke Shimizu*}

\affil[1]{Department of Engineering Science, Graduate School of Informatics and Engineering,\\
 The University of Electro-Communications, 1-5-1 Chofugaoka, Chofu, Tokyo 182-8585, Japan\\}
\affil[2]{Department of Applied Physics, Graduate School of Advanced Engineering,
Tokyo University of Science, 6-3-1 Niijuku, Katsushika-ku, Tokyo 125–8585, Japan}
\affil[*]{r-simizu@uec.ac.jp}
\date{\empty}

\begin{abstract}
Distinguishing photon-arrival time and position is crucial for advancing quantum technology. However, capturing spatial and temporal information efficiently remains challenging. Here, we present a novel photon-detection technique to achieve a significantly more efficient measurement of frequency-entangled biphoton than conventional photon detectors. We utilize a delay-line-anode single-photon detector (DLD), which consists of a position-sensitive delay line anode sensor behind a microchannel plate. Biphotons are obtained from the decay of biexcitons in the copper chloride semiconductor crystal. Two DLDs are coupled with a grating spectrometer exit to measure the joint spectral distributions of the biphoton. The resulting non-scanning process requires only a few minutes to obtain a temporally and spectrally resolved image, which is much quicker than the conventional biphoton frequency measurement. 
Our technique paves the way for all experiments in multi-mode quantum science requiring coincidence measurement.
\end{abstract}

\begin{document}

\maketitle

\section{Introduction}
Related studies on quantum optics have led cutting-edge scientific research from fundamental to practical applications. Notably, the progress of photon detection technologies has significantly underpinned the successful development of the quantum optics field. To date, the quantum efficiency of photon detectors approaches nearly 100 \% \cite{Reddy:20}, and photon-number-resolving ability is also available\cite{Cheng2022A1P}. Moreover, the detectable wavelength of photons extends to the 10 \textmu m \cite{Verma2021SinglephotonDI}, and distinguishing photon arrival timing have sub-3 ps\cite{Korzh_2020}. Similarly, the spatial resolution of photon detectors is vital to imaging or spectroscopic applications in the quantum optics but remains insufficient for the practical use of quantum optical experiments. 
\newline
\indent
Several approaches have been proposed to develop spatially resolving single-photon detectors.  
Electron Multiplying Charge-Coupled Devices (EMCCD) have been widely applied to quantum imaging\cite{doi:10.1038/s41598-018-26144-7,Bolduc_2017}.
In practice, spatiotemporal Hong-Ou-Mandel (HOM) interference observation has been reported\cite{PhysRevA.100.013845, PhysRevX.10.031031} without scanning a position-insensitive single-photon detector.
However, the frame rate considerably restricts the acceptable number of events for EMCCD; therefore, they cannot execute time-stamping measurements and more than 10,000 images are needed to reconstruct a spatiotemporal image\cite{PhysRevLett.120.203604}. 
Single-Photon Avalanche Diode (SPAD) array detectors have been a rapidly developed technology and might be a promising candidate for future quantum technologies.
A prototype 3.2 megapixel SPAD has achieved 100 ps temporal resolution and zero read out noise\cite{9720605}, and a small number of pixel SPAD has been practically applied to quantum imaging experiments\cite{Gili:23}.
Nonetheless, the current SPAD array has, in general, low fill factors and several micrometer pixel pitches for quantum optical experiments such as microscopy.
Cutting-edge superconducting nanowire single-photon detector (SNSPD) technologies have attracted great interest in the last two decades, and superconducting nanowire single-photon imager (SNSPI) structures have been evolving with several architectures, such as fiber-coupled multi-pixel arrays or single flux quantum logic interfaces\cite{doi:10.1063/5.0045990,doi:10.1021/acsphotonics.0c00433,Takeuchi:20}.
Notably, the thermally-couple imager (TCI) approach\cite{TCI_first} attained 400,000 pixels single-photon imager\cite{TCI_second}.
Yet, the SNSPD array suffers from defective pixels. Additionally, the embedding of cryogenic cooling into an imaging or spectroscopic setup to accompany the changes of other optics elements such as lenses or dispersive mediums present a challenge. 
\begin{figure*}[t]
\begin{center}
\includegraphics[scale=0.40]{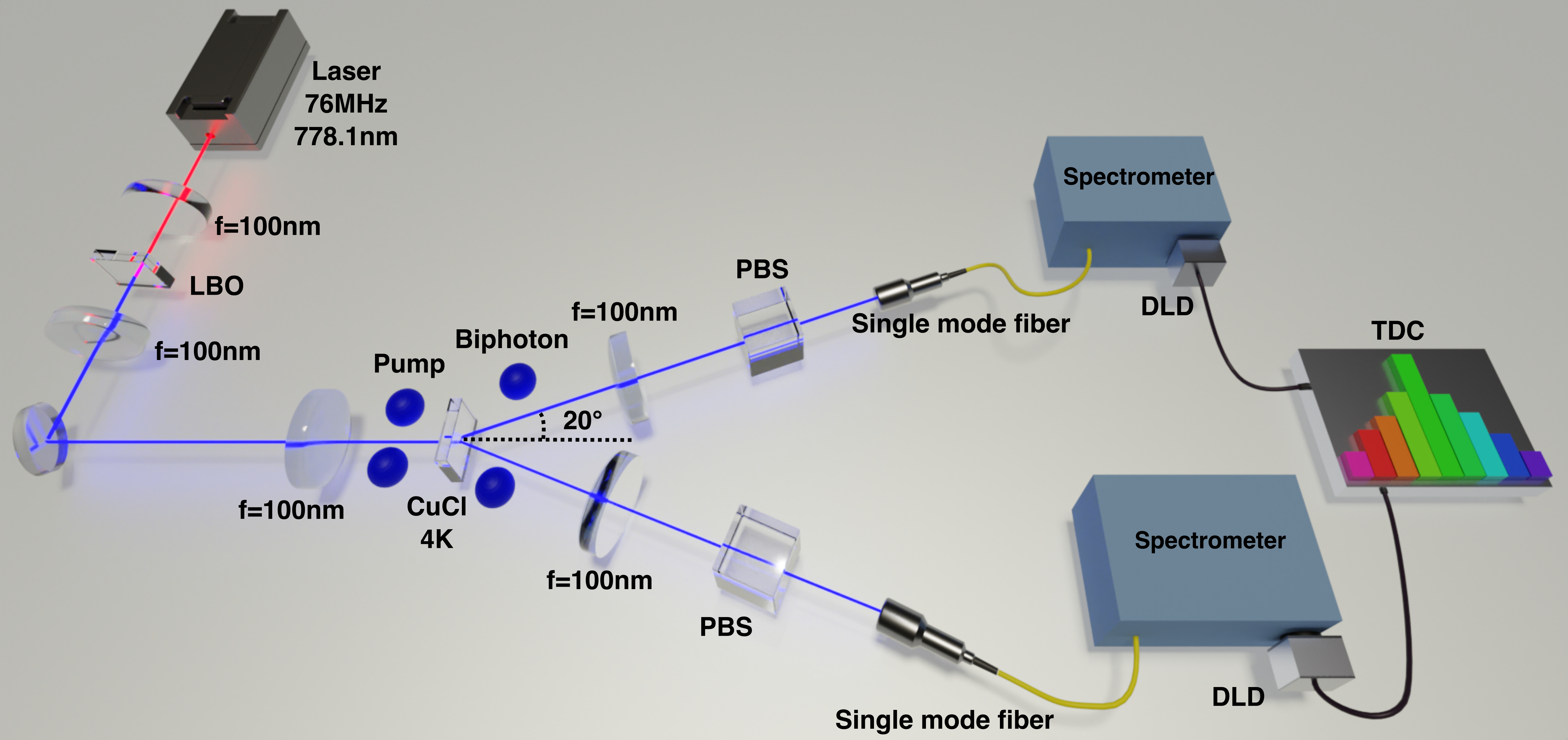}
\caption{The optical diagram for biphotons generation and detection. CuCl semiconductor single crystal is kept at 4 K and excited by second-harmonic light at 389 nm. Pump photons are filtered with a polarizing beam splitter (PBS), and scattered biphotons are collected at an angle of $20^{\circ}$ and coupled into the single-mode fibers. Coupled photons are incident into grating spectrometers and impinged on delay-line-anode single-photon detectors (DLDs). The electrical pulses are analyzed on a Time-to-Digital converter (TDC).}
 \label{Optical_diagram.png}
 \end{center}
\end{figure*}
\newline
\indent
Another promising approach is to employ position-sensitive anode sensors\cite{DLD_spectrometer} combined with a microchannel plate (MCP).
Such detectors have been developed initially to detect charged particled, but in recent years, have also been designed as a photon detector in combination with a photocathode.
For instance, a quadrant anode sensor can locate the photon-landed coordinate by splitting the detected charges toward four quadrant arcs and calculating the ratio of the charges between the four sections\cite{quadrant_anode_detector}.
A delay line anode which identifies the photon arrival position by converting the timing information into the delay line surface coordinate\cite{JAGUTZKI2002244} is another approach.
This type of detector has been applied to a wide range of scientific fields, such as fluorescence lifetime imaging microscopy\cite{SUHLING2019162365, doi:10.1063/1.4962864}, position-sensitive time-of-flight spectroscopy\cite{ARNOLD201453, DAMM201574}, and radio frequency timers\cite{MARGARYAN2022166926}.
The delay-line-anode detector (DLD) can also detect photon signals on a picosecond scale with spatial resolution. Yet, there have been no reports on quantum optical research because the photocathode sensitivity is insufficient for visible to infrared wavelength photons frequently produced via spontaneous parametric down-conversion (SPDC).
\newline
\indent 
Here, we present a temporally-resolved biphoton spectrometer with two position-sensitive DLDs to unveil the capability of the DLD for quantum optical experiments.
Biphoton spectral mapping in 2D frequency space, generally known as joint spectral intensity (JSI), is often used to clarify spectral uncorrelation \cite{article_correlation} or manipulate a biphoton spectrum \cite{10.1063/5.0059895} toward quantum communication applications. In most of those works, biphoton spectra were observed at telecom wavelengths with an optical fiber spectrometer. However, using the fiber spectrometer in the visible wavelength range presents a challenge due to significant optical loss\cite{Saleh:1084451}. Combining DLD with a grating spectrometer would mitigate the observable wavelength restrictions. Additionally, the biphotons used in most works were generated via nonlinear optical processes of spontaneous parametric down-conversion or four-wave mixing in waveguide materials. Those non-resonant processes provide a desirable biphoton state but might lack spectroscopic interest in light-matter interactions. Instead, we observe biphotons produced via biexciton in a CuCl semiconductor single crystal and show the capability of our biphoton spectrometer. Spectra of the biphotons produced via a biexciton in CuCl \cite{edamatsu2004generation} peak around 389 nm, making it suitable for the desired photocathode sensitivity. 

\section{Method}
\begin{figure*}[h]
\begin{center}
\includegraphics[scale=0.9]{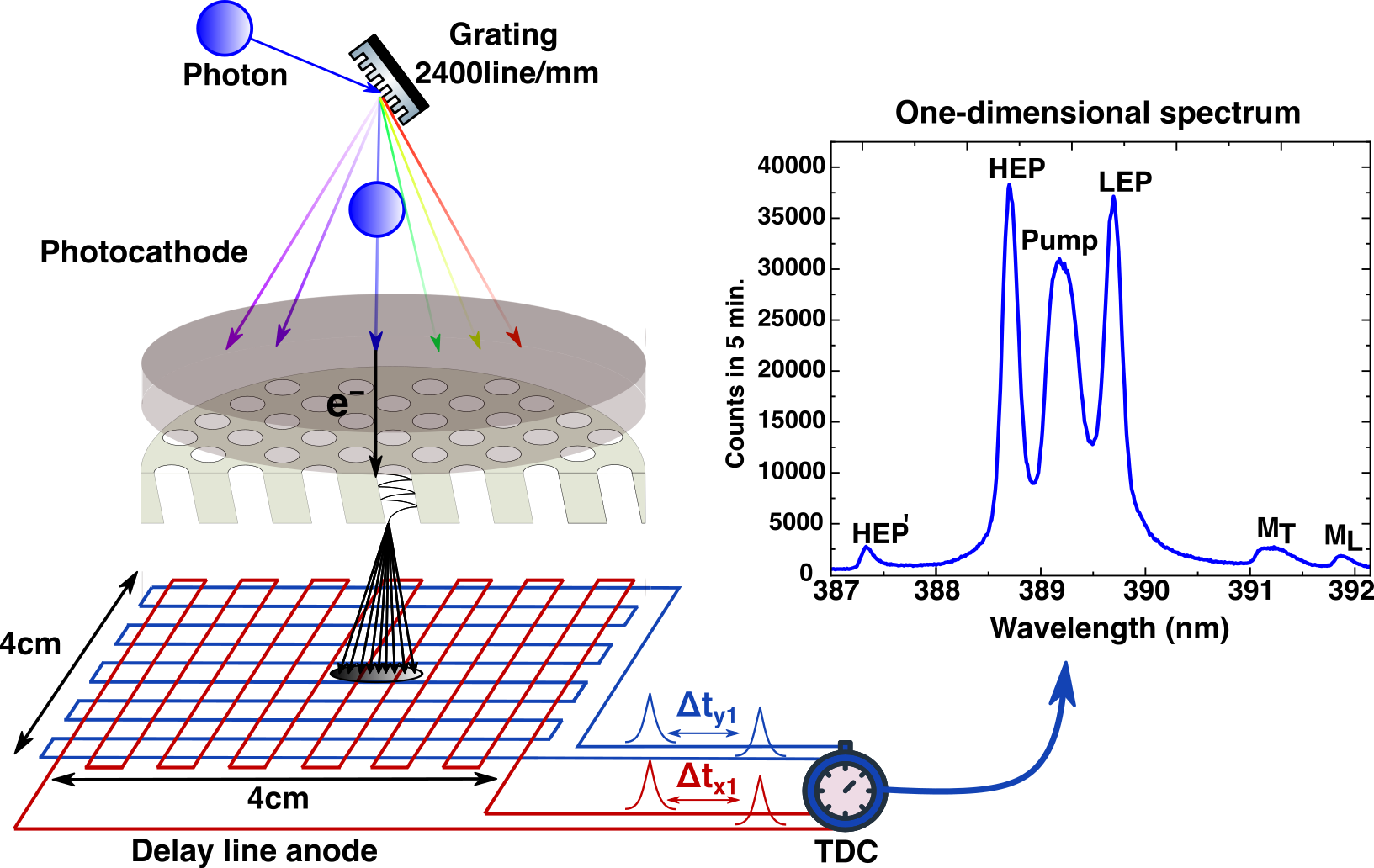}
\caption{One-dimensional spectrum measurement with one DLD. A photon is spectrally dispersed by grating optics and detected at the DLD. The incident photon is converted into a photoelectron at the photocathode and amplified through a microchannel plate followed by the delay line anode noted as blue and red on which detected pulses split into four directions. Time-to-Digital converter (TDC) acquires the timing information and constructs a one-dimensional spectrum.}
 \label{oneDLD_spectrum.png}
 \end{center}
\end{figure*}
The sample-fabrication process is detailed in our previous work \cite{PhysRevA.106.063716}.
CuCl is the first-found semiconductor material emitting polarization-entangled photons via a biexciton in a resonant hyperparametric scattering (RHPS) process\cite{edamatsu2004generation}.
This process is in conjunction with the phase-matching condition in the transition between the biexciton and exciton-polariton bands: a biexciton is produced by resonant two-photon excitation and coherently decades into two exciton-polaritons. The two polaritons scattered with the high and low energy combination, high energy polarition (HEP) and low energy polarition (LEP), respectively. 
Consequently, polaritons propagate through the crystal in the direction imposed by the phase-matching condition and transform into high-energy and low-energy photon pairs at the crystal end.
The biexciton is efficiently created by a resonant two-photon excitation by a pulsed laser source with a center wavelength of around 389 nm.
In this work, we generated the pump light using a frequency-doubled mode-locked Ti-sapphire laser which operates at around 778.1 nm with a spectral width of 0.18nm. The CuCl was installed in a cryostat with liquid helium to prevent thermal disturbance, keeping the ambient temperature at 4K. The biphoton created via RHPS is expressed in the frequency mode as
\begin{equation}
\ket{\psi} = \frac{1}{\sqrt{2}}\left(\ket{\omega_{\text{HEP}},\omega_{\text{LEP}}}
+ \ket{\omega_{\text{LEP}},\omega_{\text{HEP}}}\right),
\label{entangled_state}
\end{equation}
where $\omega_{\text{HEP}}$ and $\omega_{\text{LEP}}$ are the frequency of the high-energy polariton and low-energy polariton, respectively. The biphoton retains the entangled properties in polarization mode. However, we focus only on the frequency mode; the only spectrometer is positioned to reveal the emission spectrum without polarization analysis elements such as a half-wave plate or quarter-wave plate. 
\newline
\indent
Figure \ref{Optical_diagram.png} shows the optical diagram for biphoton generation and detection. The excitation source is prepared via second-harmonic light from a 5 mm-long type-I LBO crystal. The excitation source has a 76 MHz repetition rate. The excited crystal emits biphotons in a specific direction that is constrained by the phase-matching condition. We chose the direction to be $20^{\circ}$ from the excitation laser path to optimize the biphoton collection efficiency in this setup. These scattered biphotons are then collimated behind the CuCl sample and coupled into single-mode fibers. Polarizing beamsplitters (PBS) are positioned to exclude excitation sources from biphoton detection, thus allowing biphoton to transmit and be coupled into the fibers. These coupled photons are spectrally resolved in a spectrometer. After being separated by frequencies, each photon is detected on DLD (RoentDek), followed by a Time-to-Digital Converter (TDC, RoentDek cTDCx) to analyze timing information and map a JSI.
\newline
\indent
The DLD consists of a photocathode, a microchannel plate (MCP), and a delay-line-a sensor which perform the photon-to-photoelectron conversion, photoelectron amplification, and position specification, respectively. The flow through the photocathode and the MCP experiences a physical process as an Intensified Coupled Charged Device (ICCD). However, because the processes cannot yield spatial resolution to distinguish the position of incident photons, the delay-line-anode sensor can provide spatial resolution. A fundamental principle of the sensor is the time-to-position conversion, which operates by obtaining the arrival time of the signals propagating through a meandered wire and converting it into photon detected position\cite{JAGUTZKI2002244}. A conceptually similar architecture is employed in a superconducting-nanowire delay line single-photon imager\cite{Delayed_SNSPD}. Since the DLD possesses a high spatial and temporal resolution, the DLD has a potential application for the measurements that require spatiotemporal data; for instance, it enables time-resolved fluorescent spectral measurements\cite{SUHLING2019162365}. Additionally, combining two DLDs enables the measurement of a second-order correlation function $\text{g}^{(2)}$  because extracting and correlating signals from two MCPs is identical to the Hanbury-Brown Twiss setup\cite{doi:10.1080/14786440708520475}. 
In this paper, we combine two DLDs with a spectrometer to build a temporally-resolved biphoton spectrometer, which can execute the second-order correlation measurements and JSI mapping.
\newline
\indent
\begin{figure*}[h]
\begin{center}
\includegraphics[scale=0.80]{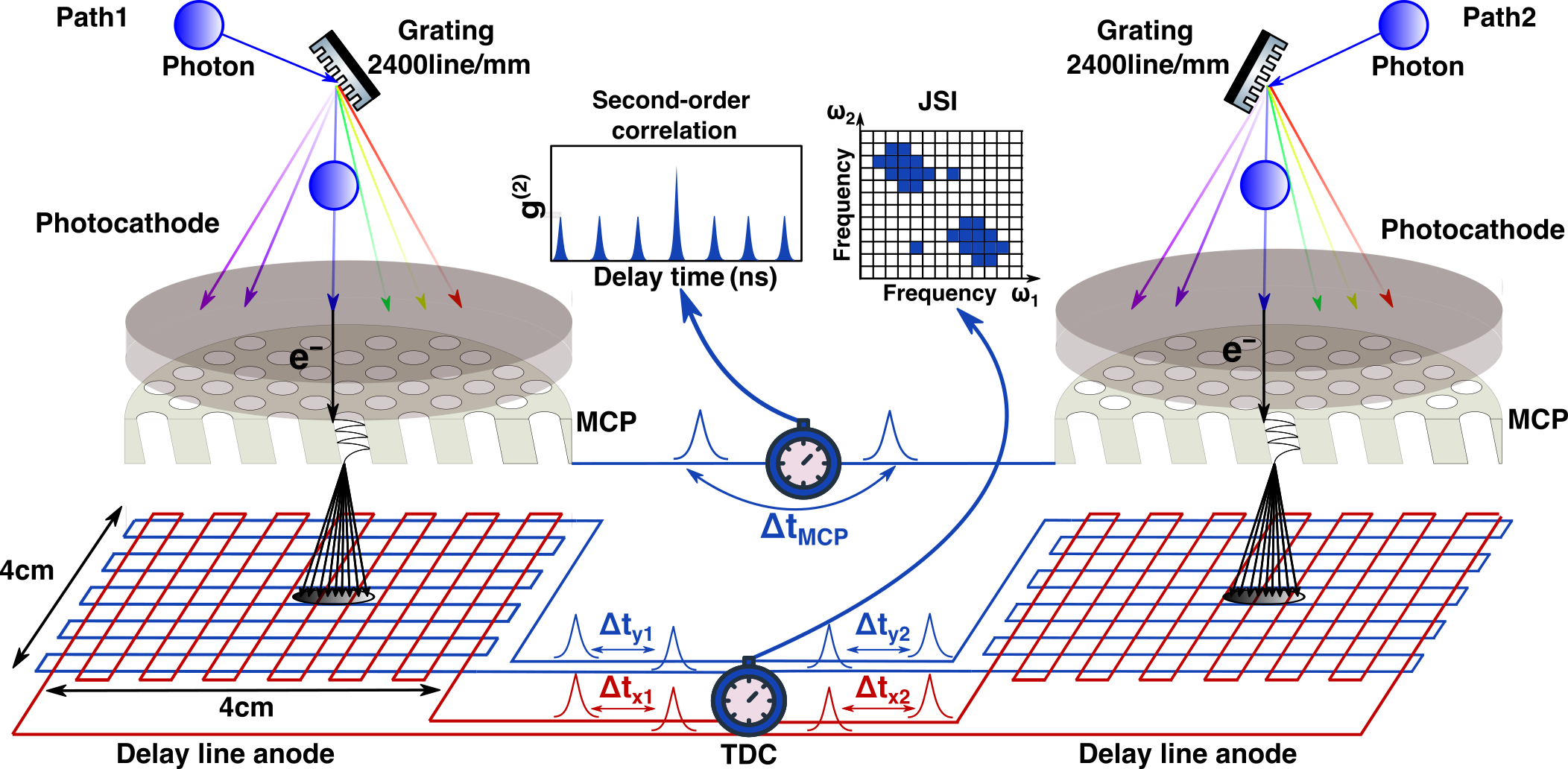}
\caption{Schematic of temporally-resolved biphoton spectrometer containing two DLDs. The second-order time correlation and Joint Spectrum Intensity (JSI) are synchronized to plot only biphoton counts in two-dimensional frequency space.}
 \label{172053_23Feb23}
 \end{center}
\end{figure*}
Before explaining the complicated biphoton measurement system used in this study, we first discuss a one-dimensional spectrum scheme containing one DLD, as shown in Fig. \ref{oneDLD_spectrum.png}. A photon is spectrally dispersed on a diffraction grating with 2400 lines/mm in a spectrometer and incident into the active photocathode area of the DLD. The dispersed photon is converted into a photoelectron through a photocathode, followed by amplification via MCP. The MCP amplifies an incident photoelectron up to $10^{6}$ times through applied high voltage at 3kV. The amplified photoelectrons land at the $4 \times 4$ $\mathrm{cm}^{2}$ delay-line-anode sensor, where two meandered lines are wired orthogonally to each other. The anode signal splits into four directions, causing it to reach the TDC at different times. Here, the time difference in the x- and y- direction between the two signals is denoted as $\Delta t_{x1}$ and $\Delta t_{y1}$, respectively. Notably, the TDC can only know the arrival time of signals. Still, it can specify photon arrival position owing to the corresponding relation between the time differences and arrival position. As a result, the single photon spectrum plotting photon count distribution as a function of frequency or wavelength can be obtained.
The right part of Fig. \ref{oneDLD_spectrum.png} shows an example of a measured one-dimensional spectrum in RHPS of a CuCl semiconductor single crystal. The spectrum shows the HEP peak at around 388.8 nm and the LEP peak at around 389.8 nm. The peak of the Rayleigh-scattered pump source appears around 389.2 nm, sandwiched by HEP and LEP peaks. While these high three peaks are observed in the range of 388.5 to 390 nm, the small side tails can be seen labeled as $\text{HEP}^{\prime},\text{M}_{\text{T}}$, and $\text{M}_{\text{L}}$. The three peaks originate from biexciton luminescence\cite{PhysRevA.106.063716}; they do not exhibit a biphoton state and are much smaller than the prominent three peaks. Hence, we ignore their counts on the following discussions, given that their count contribution is negligibly small compared to the major three peaks.
\newline
\indent
We expand the single DLD system into a two-DLD system for the JSI measurement. Figure \ref{172053_23Feb23} illustrates the schematic of the temporally-resolved biphoton spectroscopic system containing two DLDs. Here, it is assumed that time-correlated biphotons are incident onto individual DLDs. The functions of the diffraction grating, DLD, and TDC are identical to those in the case of a single-photon spectrum. Contrary, the spectrum is changed from one-dimensional to two-dimensional mapping. The two-dimensional mapping, a JSI, is constructed by plotting coincident biphotons from each DLD. Unnecessary photon counts such as pump photons, however, blur the contrast of the JSI. Such unpreferable counts should therefore be excluded by time-correlation measurement. Electrical signals guided by the MCPs can execute the time-correlation measurement or a second-order correlation function $\text{g}^{(2)}$ analysis. The second-order correlation function generally shows a higher center peak at the zero-delay time when the JSI is measured, indicating that biphotons are simultaneously scattered from biexcitons in CuCl. These time-correlated counts are an indicator that only biphoton counts are being plotted on a spectrum. Hence, we first investigated a second-order correlation function to verify the incident timing of the biphoton. After the timing verification, we mapped observed biphoton counts that peaked at a zero-delay time onto 2D frequency space.  
\newline
\indent 
The time-to-position conversion can be described using straightforward linear equations. 
To translate the raw timing information from four lines into the photon detection position, the time difference is calculated on each anode, noted as $\Delta t_{x1},\Delta t_{y1},\Delta t_{x2}$, and $\Delta t_{y2}$.   These time lags are incorporated into Eq. (\ref{coordinate1}) and (\ref{corrdinate2}) to identify the positions.
\begin{equation}
    x_{1}(\Delta t_{x_{1}}) = \frac{(\Delta t_{x_{1}} + t_{a})v}{2},\quad
    y_{1}(\Delta t_{y_{1}}) = \frac{(\Delta t_{y_{1}} + t_{a})v}{2} \label{coordinate1}
\end{equation}
\begin{equation}
    x_{2}(\Delta t_{x_{2}}) = \frac{(\Delta t_{x_{2}} + t_{a})v}{2},\quad
    y_{2}(\Delta t_{y_{2}}) = \frac{(\Delta t_{y_{2}} + t_{a})v}{2}
    \label{corrdinate2}
\end{equation}
Here, $v$ is the speed at which the electrical signals propagate on the wires, and $t_{a}$ is the time in which the signals propagate from the edge to the entire delay-line-anode. The coordinates $(x_{1},y_{1})$ and $(x_{2},y_{2})$ represent the photon detection position on each DLD.  In this work, the photon detection position is transformed from position to frequency space through the spectrometers.
On the biphoton spectrometer, we plot a coincidence count as a function of two-dimensional coordinates only when the biphotons simultaneously trigger the two DLDs. Consequently, two-dimensional photon mapping in frequency space is plotted, where the x- and y-axis represent the frequency of a photon detected at path 1 and path 2, respectively. 
\section{Results and discussion}
\begin{figure*}[h]
\begin{center}
\includegraphics[scale=0.35]{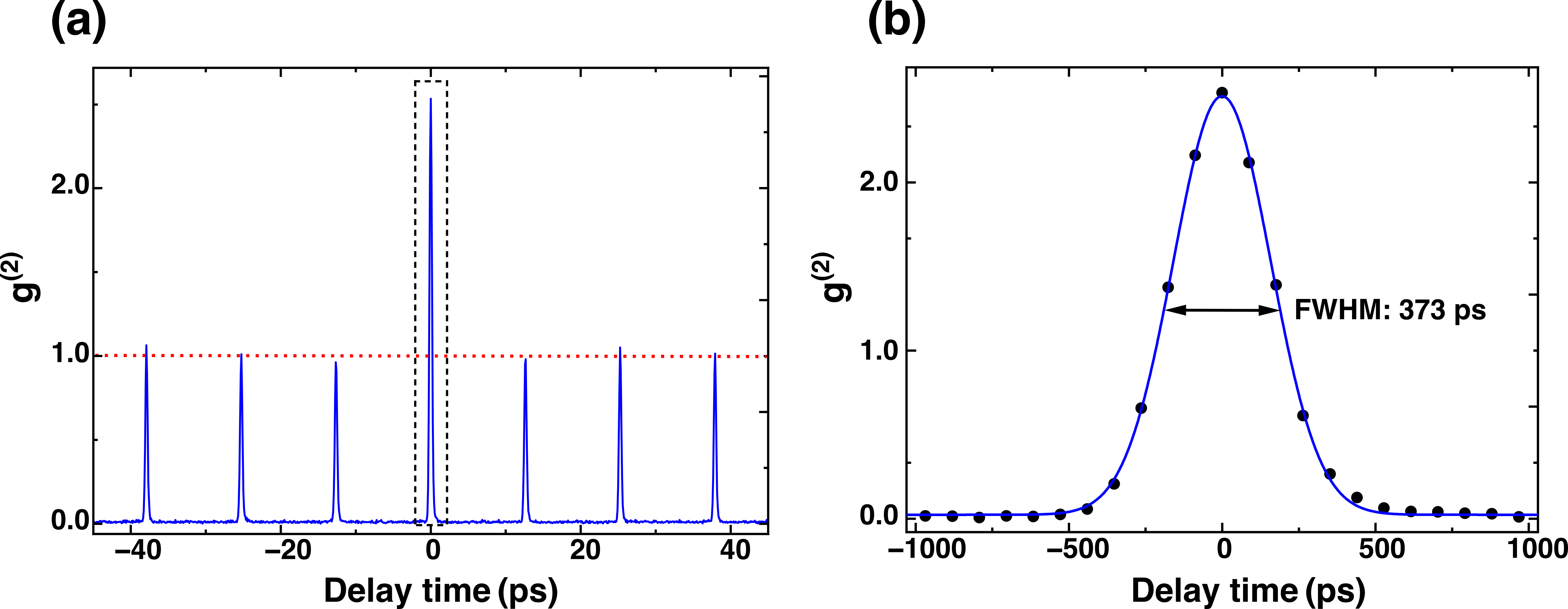}
\caption{(a) Second-order correlation function as a function of the delay time between two MCPs. The bin size is 88 ps. (b) Center peak showing the inside area of the dotted line in (a).}
 \label{171501_23Feb23}
\end{center}
\end{figure*}

\begin{figure}[H]
\begin{center}
\includegraphics[scale=0.50]{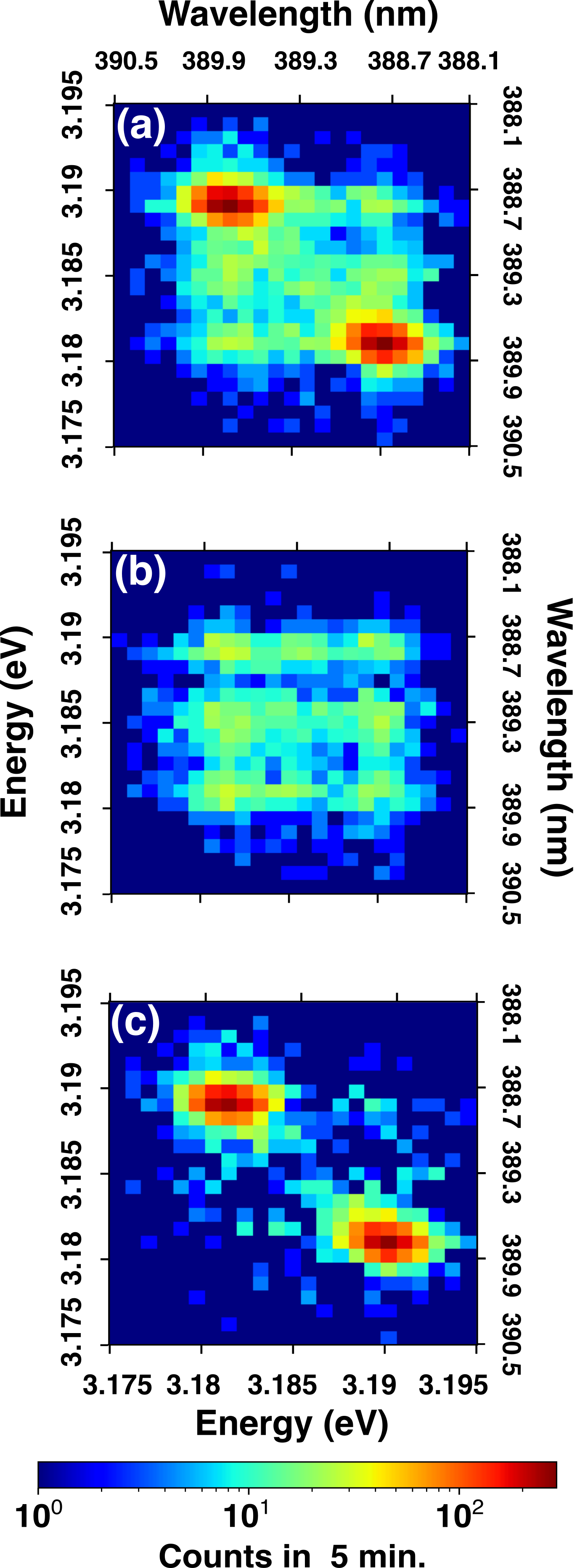}
\caption{Biphoton spectra at two DLDs. (a) The counts including $\text{HEP}, \text{Pump}, \text{LEP}$
(b) The accidental counts including only one side peak at 13 ns in Fig. \ref{171501_23Feb23} (c)The counts after subtracting accidental counts.}
 \label{235122_28May23} 
\end{center}
\end{figure}

Figure \ref{171501_23Feb23} (a) shows the second-order correlation function, correlated between two MCPs counts embedded inside the DLDs. As discussed in our previous work \cite{PhysRevA.106.063716}, the center peak at the zero-delay time, surrounded by dotted lines, is nearly two times higher than the side peaks averaged on 1.0 over all side peaks. This center peak indicates that biphotons simultaneously impinged at two DLDs because the biphotons are coincidentally scattered from the CuCl semiconductor crystal. The small side peaks appeared at every 13.2 ns, a reciprocal of the laser repetition rate of 76 MHz. The side peaks are considered to be mainly originating from Rayleigh scattered pump photons from the CuCl sample. It should be noted that the center peak at zero-delay time contains biphotons and accidental coincidence counts. We define the counts at the center peak ranging from -500 ps to 500 ps as coincidence counts.
\newline
\indent
Figure \ref{171501_23Feb23} (b) shows the enlarged plots of Fig. \ref{171501_23Feb23} (a) around the zero-delay position. The peak was fitted to a Gaussian function, and full-width half maximum (FWHM) was calculated to estimate the temporal resolution of the detection system. The FWHM was 373 ps, leading to the 263 ps temporal resolution per DLD, given that the observed distribution is provided by the convolution of the signal from two identical DLDs. The 263 ps timing resolution is lower than that of the other single-pixel single-photon detectors, sub-3 ps at SNSPD or 50 ps at SPAD, but superior to the other single-photon imagers such as an arrayed SNSPD or SPAD array. EMCCD detectors can reconstruct coincidence images as a single-photon imager, yet they cannot resolve photon counts in a picosecond scale\cite{PhysRevLett.120.203604} as the DLD does. 
\newline
\indent
Figure \ref{235122_28May23} illustrates biphoton spectra where the horizontal (vertical) axis represents the energy for path 1 (path 2). The corresponding wavelength is added on the top and right axes. The accumulation time is 5 mins, with the coincidence of biphoton counts plotted on a logarithmic scale. The two vivid peaks appear on the upper left and lower right, indicating that coincidence counts are dominantly distributed with higher and lower energy pairs. The distribution is consistent with the phase-matching condition that two pump photons are converted into a lower and higher energy biphoton pair via RHPS, as expressed in Eq. (\ref{entangled_state}). However, the subtle background counts caused by pump source photons still remain present near the two peaks. This is mainly because the pump counts are also piled at the center peak on the coincidence histogram in Fig. \ref{171501_23Feb23}. Here, the coincidence count on the upper left is 289, and the pump count maximum is 27; therefore, the coincidence-to-accidental ratio (CAR) is 10.7. 
\newline
\indent
The JSI of the coincidence histogram at 13.2 ns in Fig. \ref{171501_23Feb23} (a) was plotted in Fig. \ref{235122_28May23} (b), showing that a spectrum of accidental coincidence counts is mainly distributed across all nine peaks. The cause of these nine peaks is accidental coincidence counts with the combination of the HEP, LEP, and pump peak spectra. To enhance the CAR on the JSI, Fig. \ref{235122_28May23} (b) is subtracted from Fig. \ref{235122_28May23} (a). Here, all accidental count peaks are assumed to possess the same spectral distribution; plotting another side peak yields a similar spectrum distribution. The result confirms that the pump source counts hide behind the center peak.
The subtracted spectrum is shown in Fig. \ref{235122_28May23} (c), vividly and sharply showing only two biphoton peaks. The upper left counts peak at 259, and the signal maximum is 1, therefore, yielding a CAR of 259. Compared with Fig. \ref{235122_28May23} (a), an improvement of over 25 times is observed in the CAR. Hence, the subtraction procedure also functions as a further filtering role in addition to the coincidence windows. Nevertheless, tiny background counts, for instance, between two biphoton peaks or above the lower right peak, still exist and originate from other possible accidental counts such as $\text{HEP}^{\prime}$, $\text{M}_{\text{T}}$ counts, or dark counts generated inside the 
photocathode.
Several experimental conditions could be considered; for instance, the multi-hit events were wholly ignored in this setup which reduced the measurement efficiency. The meander-shaped delay line anode sensor plays a crucial role in identifying the photon-arrival position. This condition indicates that the measured spectra do not contain all photons and are, therefore, expected to contain lower counts than the actual arrived photon distribution. 
The most direct and easy-to-change alternative is to utilize a hexagonal delay line anode to distinguish the multi-hit events at the relative blind zone\cite{1046770}. On the hexagonal delay line anode, three delay line anode wires are crossed by 60 $^{\circ}$ to prevent signals from propagating in a horizontal and vertical direction, suppressing the dead time of less than 10 ns. Although this hexanode delay line anode is superior to the square delay line anode used here, it cannot process an event in which the two photons are detected simultaneously at the same position.
\newline
\indent 
It is also worth mentioning that this technique essentially omits information on one of the two axes from each detector's image, only extracting the x-projected count distribution from each detector and calculating the joint distribution between them. More generally, combining two delay line anode sensors can acquire $2\times 2 = 4$ dimensional count distribution, yet preceding processes until sensors are unsuitable for maximizing 4-dimensional information. Positioning the delay line anode sensor behind the streak tube\cite{hamamatsu-manual} would be a promising approach to address these issues. A streak camera sweeps the photoelectrons generated by the photocathode and constructs an image on a phosphor screen where the horizontal and vertical projections contain spatial and temporal information, respectively. Replacing the phosphor sensor with a delay line anode sensor and computing the correlation between two detectors would enable the computation of spectrally- and temporally- resolved biphoton distribution and enable us to map biphoton distribution in time- and frequency-domain simultaneously\cite{10.1063/5.0059895}.
\newline
\indent
Biphoton wavelength sensitivity matches around at approximately 388 nm, and the photon count loss is relatively low. The photocathode sensitivity of our DLD peaks at around the UV region, whose quantum efficiency is approximately 20\% at 300 nm and nearly 0\% at 800 nm. Hence, it is still impossible to use DLDs with the photocathode in this paper for the near-infrared imaging or spectroscopy with biphoton produced from spontaneous parametric down-conversion. However, such measurements should be executable by replacing them with a visible-sensitive photocathode and keeping the temperature low enough to suppress thermal noises. \newline \indent
The DLD-based system would have broad applications in quantum optical science. For example, replacing the grating spectrometer with a simple imaging system, quantum ghost imaging\cite{spatial_en}, or quantum microscopy\cite{quantum_metrology} would be possible. Additionally, investigating the wavelength-dependence of single-photon absorption and emission is possible by combining DLDs with fluorophore such as a natural photosynthetic complex\cite{phtosynthetic_complex}. The temporally resolved biphoton spectrum while synchronizing with second-order correlation function \cite{doi:10.1021/acsphotonics.2c00817} is also plausible with our system, for example, the spectrally and picosecond time-resolved antibunching measurement\cite{psTRAB}. More generally, all quantum metrology with $g^{2}$ measurement and spectral- or spatial mapping will be executable by a DLD-based measurement system. 
\section{Conclusion}
We demonstrated a spectrally-temporally-resolved biphoton detection technique with delay-line-anode detectors. Firstly, we observed a one-dimensional spectrum with only one DLD. Moving onto two-DLDs measurement, picosecond coincidence measurement between microchannel plates was successfully measured, serving as a pump photon counts filtration for 2D spectra. Based on the time-resolved histogram, 2D spectra originating from only biphotons were observed without a time-wasting scanning process via spectrometer or piling the gated images as EMCCD does. In addition, the technique presented in this paper is helpful for JSI measurement and all coincidence measurements on quantum optics research requiring spectrally- or spatially- resolving capability: quantum spectroscopy, microscopy, imaging, and communication.

\begin{backmatter}
\bmsection{Funding}
This work was supported by MEXT Quantum Leap Flagship Program (MEXT Q-LEAP) Grant No.JPMXS0118069242.
\end{backmatter}

\bibliography{reference}
\end{document}